\documentclass[aps,prl,twocolumn,groupedaddress]{revtex4-1}
\usepackage{amsmath}
\usepackage{amssymb}
\usepackage{graphicx}
\usepackage{dcolumn}
\usepackage{bm}

\begin{document}

\title{Band dynamics accompanied by bound states in the continuum at the third-order $\Gamma$ point in leaky-mode photonic lattices}
%\title{Dirac cone dispersions accompanied by extended bound states in the continuum at the fourth stop bands of leaky-mode photonic lattices}

\author{Sun-Goo Lee}
\email{sungooleee@gmail.com}
\author{Seong-Han Kim}
\author{Chul-Sik Kee}
\email{cskee@gist.ac.kr}
\affiliation{Integrated Optics Laboratory, Advanced Photonics Research Institute, GIST, Gwangju 61005, South Korea}
\date{\today}

\begin{abstract}
Bound states in the continuum (BICs) and Fano resonances in planar photonic lattices, including metasurfaces and photonic crystal slabs, have been studied extensively in recent years. Typically, the BICs and Fano resonances are associated with the second stop bands open at the second-order Gamma ($\Gamma$) point. This paper address the fundamental properties of the fourth stop band accompanied by BICs at the third-order $\Gamma$ point in one-dimensional (1D) leaky-mode photonic lattices. At the fourth stop band, one band edge mode suffers radiation loss, thereby generating a Fano resonance, while the other band edge mode becomes a nonleaky BIC. The fourth stop band is controlled primarily by the Bragg processes associated with the first, second, and fourth Fourier harmonic components of the periodic dielectric constant modulation. The interplay between these three major processes closes the fourth band gap and induces a band flip whereby the leaky and BIC edges transit across the fourth band gap. At the fourth stop band, a new type of BIC is formed owing to the destructive interplay between the first and second Fourier harmonics. When the fourth band gap closes with strongly enhanced $Q$ factors, Dirac cone dispersions can appear at the third-order $\Gamma$ point. Our results demonstrate a method for manipulating electromagnetic waves by utilizing the high-$Q$ Bloch modes at the fourth stop band.
\end{abstract}


\maketitle

%\section{Introduction}
Subwavelength planar photonic lattices, such as metasurfaces \cite{Kildishev2013,NYu2014,SSun2019,Arbabi2015,AMHWong2018} and photonic crystal slabs \cite{Joannopoulos1995,Johnson1999}, are used extensively to manipulate electromagnetic waves. Phase-matched Bloch modes with finite $Q$ factors in the radiation continuum enable incident light to be absorbed and reemitted at a resonant frequency \cite{YHKo2018}. Based on the Fano resonances, or guided-mode resonances, various optical devices, such as reflectors \cite{Magnusson2014,JWYoon2015}, filters \cite{WShu2003,MNiraula2015,Kawanishi2020}, polarizers \cite{KLee2014,Hemmati2019}, lasers \cite{Kodigala2017,STHa2018}, and sensors \cite{YLiu2017,Abdallah2019} have been implemented to date. However, under appropriate conditions, the Bloch mode in the radiation continuum is completely decoupled from radiating waves and becomes a bound state in the continuum (BIC), thereby representing an exceptional eigensoultion of Maxwell's equations with infinite lifetime \cite{Yang2014,Plotnik2011,Hsu2016,Koshelev2019}. Recently, robust BICs in a photonic-crystal slab geometry have received widespread attention because they are associated with fascinating and diverse physical phenomena such as topological natures \cite{BZhen2014,Doeleman2018,YXGao2017}, enhanced nonlinear effects \cite{Koshelev2020}, and sharp Fano resonances  \cite{SGLee2017,Koshelev2018,Abujetas2019}.

In one-dimensional (1D) and two-dimensional (2D) photonic lattice slabs, most of the important properties of Fano resonances and BICs are associated with the second stop bands that are open at the second-order Gamma ($\Gamma$)
 point \cite{Marinica2008,Hsu2013,SGLee2019-1,Rosenblatt1997}. With proper in-plane rotational symmetry, photonic lattices admit both leaky and nonleaky edges at the second stop bands \cite{Kazarinov1985,Ochiai2001,SFan2002}. In principle, not only the second stop bands but also the fourth bands, which open at the third-order $\Gamma$ point, can exhibit BICs and diverse zero-order spectral responses in the subwavelength regime. However, to the best of our knowledge, no detailed study on the fourth stop band in leaky-mode photonic lattices has been reported thus far. In this paper, we elucidate the fundamental properties of the fourth stop bands at the third-order $\Gamma$ point. In leaky-mode photonic lattices, the fourth band gaps are controlled primarily by the first-order diffraction due to the fourth Fourier harmonic lattice component, and to a lesser extent by the second- and fourth-order diffractions produced by the second and first Fourier harmonic components, respectively. The interplay of these three major diffraction processes closes the fourth band gap and induces a band flip whereby the leaky edge and the BIC edge transit across the fourth band gap. We show that at the fourth stop band, a new type of BIC is formed owing to the destructive interplay between the diffractions by the first and second Fourier harmonic components.

\begin{figure}[b]
\includegraphics[width=8.3 cm]{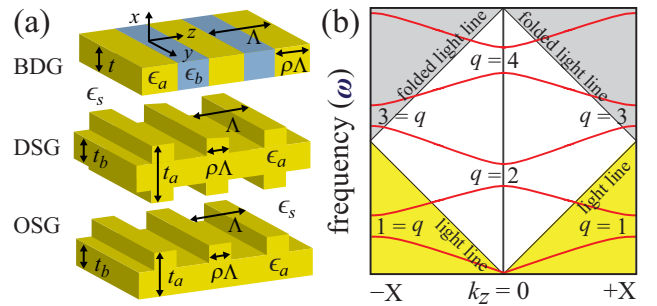}
\caption {\label{fig1} (a) Schematics of the representative 1D photonic lattices for studying the fourth stop band. (b) Conceptual illustration of the photonic band structures including first four band gaps. In general, guided waves are described by the complex frequency $\Omega = \omega - i \gamma$, where $\gamma$ represents the decay rate of the mode. }
\end{figure}

\begin{figure*}[t]
\centering
\includegraphics[width=17.0cm]{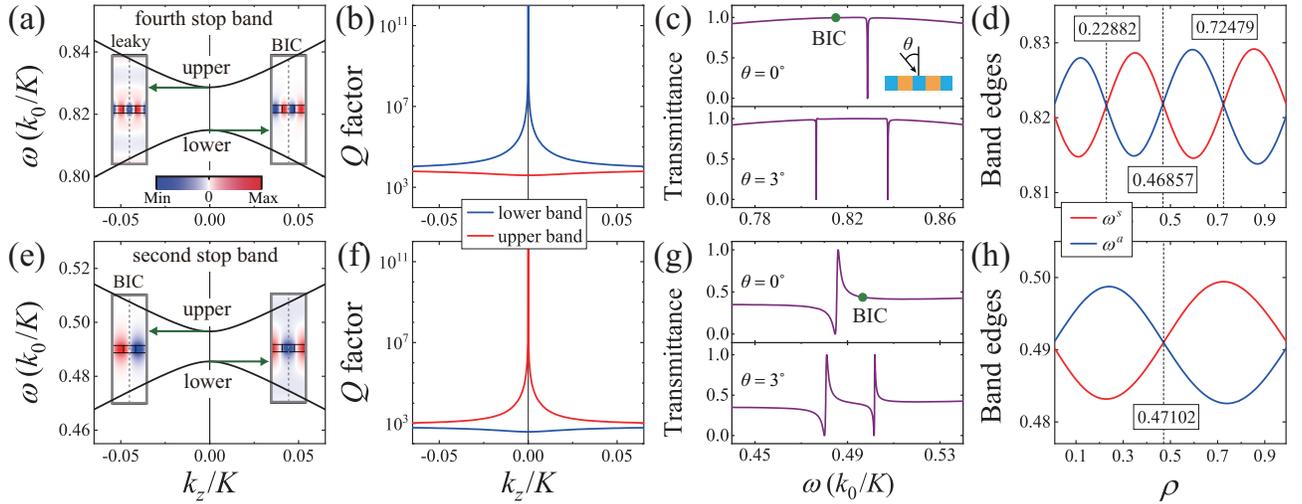}
\caption{\label{fig2} Comparison between the key properties of the fourth (a)-(d) and second (e)-(h) stop bands. (a), (e) Dispersion relations, (b), (f) radiative $Q$ factors, (c), (g) transmission spectra, and (d), (f) the evolution of the band edge frequencies as a function of $\rho$. The spatial electric field ($E_y$) distributions in the insets in Figs. (a) and (e) indicate that one of the band edge modes becomes the symmetry-protected BIC. As $\rho$ varies from 0 to 1, the fourth stop band exhibits the closed band states three times, while the second stop band shows one band gap closure. The structural parameters $\epsilon_{0}=9.00$, $\Delta\epsilon=2.00$, $\epsilon_{s}=1.00$, $t=0.20~\Lambda$, and $\rho = 0.35$ were used in the FEM simulations. }
\end{figure*}

%\section{Lattice structure and perspective}
Figure \ref{fig1}(a) depicts three representative 1D photonic lattices that support BICs and Fano resonances. The binary dielectric grating (BDG) comprises materials with high ($\epsilon_a$) and low ($\epsilon_b$) dielectric constat.
The lattice constant is $\Lambda$, the thickness of the grating layer is $t$, and the width of the high dielectric constant sections is $\rho\Lambda$. The double-sided grating (DSG) and one-sided grating (OSG) are composed of a slab waveguide layer with a grating layer attached to both and one of its interfaces, respectively. The DSG and OSG consist of dielectric material ($\epsilon_a$) with thickness of $t_a$ and $t_b$ for the thick and thin regions, respectively. As shown in Fig.~\ref{fig1}(b), owing to these periodic modulations of the lattice parameters, the representative 1D lattices exhibit photonic band gaps at the Bragg condition $k_z = qK/2$, where $k_z$ is the Bloch wavevector, $K=2 \pi/\Lambda$ is the magnitude of the grating vector, and $q$ is an integer representing the Bragg order. The fourth ($q=4$) and second ($q=2$) stop bands in the white region can be useful in practical applications because they can allow diverse zero-order spectral responses via resonant coupling with outgoing plane waves. The first stop band ($q=1$) in the yellow region is not associated with the leaky-wave effects because it is protected by total internal reflection, and the leaky third stop band ($q=3$) in the gray region is less practial because it generates unwanted higher-order diffracted waves alongside the zero-order waves \cite{YDing2007}. In general, the 1D lattices shown in Fig.~\ref{fig1}(a) can support multiple TE-polarized guided modes provided that their effective dielectric constants are larger than that of the surrounding medium ($\epsilon_{s}$), and each mode may possess multiple band gaps \cite{Magnusson2009}. In this study, we focus on the fourth stop band of the fundamental $\mathrm{TE}_{0}$ mode as it underlies the key properties of the fourth stop band. The three representative 1D lattices shown in Fig.~\ref{fig1}(a) are investigated through the rigorous finite element method (FEM) simulations.

\begin{figure*}[t]
\includegraphics[width=17 cm]{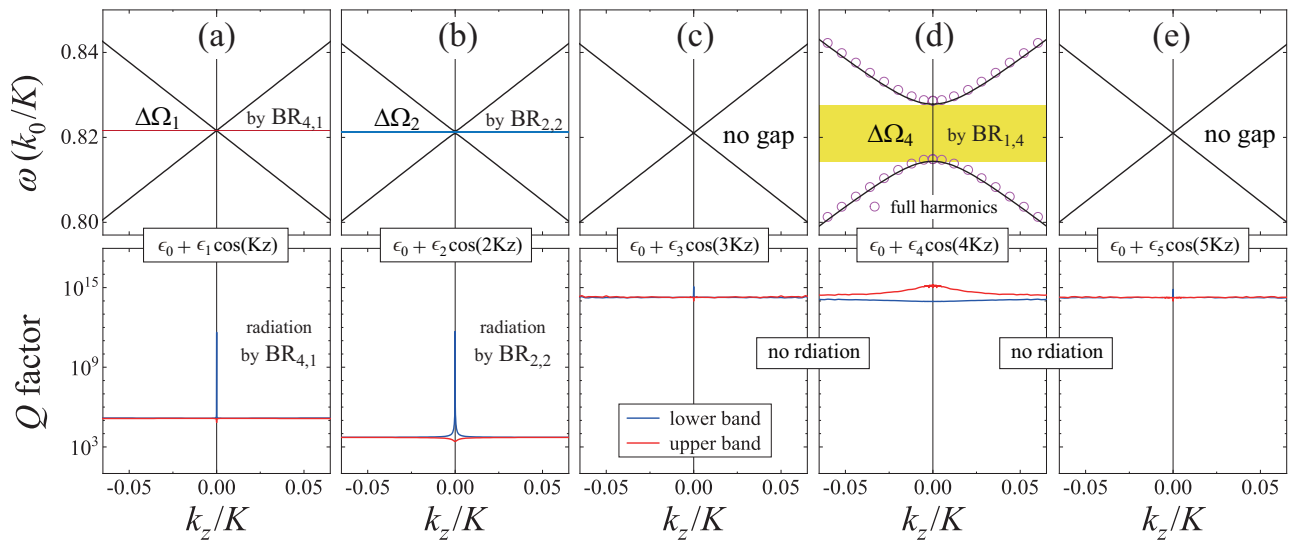}
\caption {\label{fig3} Simulated dispersion relations near the third-order $\Gamma$ point in 1D leaky-mode lattices relative to the Fourier harmonic content. The dielectric functions vary as (a) $\epsilon_{0}+\epsilon_{1}\cos(K z)$, (b) $\epsilon_{0}+\epsilon_{2}\cos(2K z)$, (c) $\epsilon_{0}+\epsilon_{3}\cos(3K z)$, (d) $\epsilon_{0}+\epsilon_{4}\cos(4K z)$, and (e) $\epsilon_{0}+\epsilon_{5}\cos(5K z)$. At the fourth stop band, out-of-plane radiation is induced by the interplay between the first and second Fourier harmonics, while the size of the gap is determined by the interplay between the first, second, and fourth Fourier harmonics. Other than the dielectric functions, the lattice parameters are the same as for Fig. \ref{fig2}.}
\end{figure*}
%\section{Leaky-band dynamics}
Key properties of the BDG structure at the fourth stop band are illustrated in Figs. \ref{fig2}(a)--\ref{fig2}(d), with equivalent plots representing the second stop band shown in Figs. \ref{fig2}(e)--\ref{fig2}(h) for comparison. As shown in Fig. \ref{fig2}(a) (Fig. \ref{fig2}(e)), the fourth (second) stop band opens at $k_z = 0$ owing to the periodic modulations in the dielectric constants. Simulated spatial electric field ($E_y$) distributions in the insets in Fig. \ref{fig2}(a) (Fig. \ref{fig2}(e)) show that the upper (lower) band edge mode with symmetric field distributions radiates from the grating layer, while the lower (upper) edge mode with asymmetric field distributions is within the BDG structure. The existence of a symmetry-protected BIC at the fourth (second) stop band can be observed clearly by investigating the radiative $Q$ factors plotted in Fig. \ref{fig2}(b) (Fig. \ref{fig2}(f)). At the fourth (second) stop band, the symmetry-protected BIC in the lower (upper) band exhibits a $Q$ factor exceeding $10^{14}$ at the third-order (second-order) $\Gamma$ point, but the $Q$ values decrease abruptly as $k_z$ moves away from the $\Gamma$ point. Next the transmission properties through the BDG with different incident angles $\theta$ in the vicinity of the fourth and second stop band were investigated. As shown in Fig. \ref{fig2}(c) (Fig. \ref{fig2}(g)), at normal incidence with $\theta = 0^\circ$, only one resonance due to the leaky mode in the upper (lower) edge of the fourth (second) stop band is observed. The symmetry-protected BIC at the lower (upper) edge of the fourth (second) stop band is not shown in the transmittance curve because it is a perfectly imbedded eigenvalue with an infinite $Q$ factor and varnishing line width. By contrast, when $\theta = 3^\circ$, two resonances due to the Bloch modes in the upper and lower bands appear simultaneously in the transmission spectra, and their locations follow the phase matching condition $k_z = k_s \sin \theta $, where $k_{s}$ represents the wave number in the surrounding medium.

The simulated results illustrated in Figs. \ref{fig2}(a)--\ref{fig2}(c) and Figs. \ref{fig2}(d)--\ref{fig2}(f) evidently reveal that the BDGs support BICs and Fano resonances at both the fourth and second stop bands, respectively. However, the relative positions of a leaky edge, $\omega^s$, and a bound state edge, $\omega^a$, are reversed for the second and fourth stop bands. Figures \ref{fig2}(d) and \ref{fig2}(h) show the evolution of the band edge frequencies $\omega^s$ and $\omega^a$ for the fourth and second stop bands, respectively, as a function of $\rho$. As shown in Fig. \ref{fig2}(h), as $\rho$ increases from zero, the second band gap opens and its size first increases, then decreases, and becomes zero when $\rho = 0.47102$ (i.e., a closed band state). The second band gap reopens and its size increases, decreases, and approaches zero when $\rho$ is increased further and approaches 1. Before and after the band gap closure, the symmetry-protected BICs are located at the upper and lower band edges, respectively. Conversely, as $\rho$ varies from zero to 1, the fourth stop band exhibits three closed band states when $\rho = 0.22822$, 0.46857, and 0.72479, as shown in Fig. \ref{fig2}(d). Before and after the band gap closures, the relative positions of the leaky and bound state edges are reversed.

In principle, the dispersion relations of the guided modes in the BDG structures can be obtained using the 1D wave equation \cite{Yariv1984}:
\begin{equation}\label{wave-equation}
\left (\frac{\partial^{2}}{\partial x^{2}}  + \frac{\partial^{2}}{\partial z^{2}} \right ) E_{y}(x,z) + \epsilon (x,z) k_{0}^2 E_{y}(x,z)= 0,
\end{equation} 	 	
where $k_{0}$ represents the wave number in free space. Equation~(\ref{wave-equation}) can be solved by expanding the periodic dielectric function $\epsilon (x,z)$ in a Fourier series and the electric field $E_{y}$ as a Bloch form \cite{Inoue2004}. In previous studies, Eq. (\ref{wave-equation}) was solved near the second-order $\Gamma$ point by employing the semi-analytical dispersion model in which the dielectric function is approximated as an even cosine function $\epsilon(z)=\sum_{0}^{2} \epsilon_{n}\cos(n\mathrm{K}z)$, where the Fourier coefficients are given by $ \epsilon_{0}=\epsilon_{l}+\rho \Delta \epsilon$ and $ \epsilon_{n\geq1}=(2\Delta \epsilon / n\pi) \sin (n\pi \rho)$ \cite{Kazarinov1985,SGLee2019-2}. The simple dispersion model indicates that the second stop band is controlled primarily by the first-order Bragg process, $\mathrm{BR}_{1,2}$, due to the second Fourier harmonic $\epsilon_{2}\cos(2K z)$, and to a lesser extent by the second-order Bragg process, $\mathrm{BR}_{2,1}$, due to the first Fourier harmonic $\epsilon_{1}\cos(K z)$. As $\rho$ varies from 0 to 1, a critical value of $\rho$ at which the second band closes because $\mathrm{BR}_{1,2}$ and $\mathrm{BR}_{2,1}$ have the same strength and opposite phase should arise; this occurs near $\rho = 0.5$ because the strength of $\mathrm{BR}_{1,2}$, which is proportional to $\epsilon_{2}=(\Delta \epsilon/\pi) \sin (2\pi \rho)$, diminishes and becomes zero as $\rho$ approaches 0.5. A band flip occurs because the primary scattering process, $\mathrm{BR}_{1,2}$, tends to locate the BIC at the lower (upper) band edge when $\rho < 0.5$ ($\rho > 0.5$). The analytical dispersion describes the simulation results for the second stop band, illustrated in Figs. \ref{fig2}(e)--\ref{fig2}(h), with impressive accuracy. At the second stop band, the out-of-plane radiation loss is caused by $\mathrm{BR}_{2,1}$.

\begin{figure}[b]
\includegraphics[width=8.3 cm]{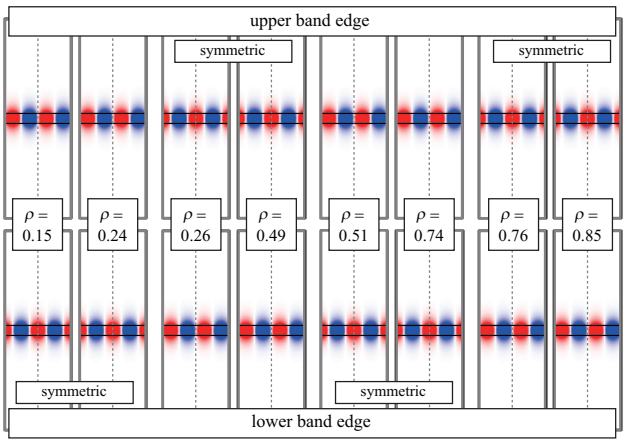}
\caption {\label{fig4} FEM simulated spatial electric field distributions at the upper and lower bands of $\Delta\Omega_4$ due to the fourth Fourier harmonic. The relative positions of the symmetric and asymmetric edges are reversed with the sign change of the fourth-order Fourier coefficient $\epsilon_4 = (\Delta \epsilon/2\pi) \sin (4\pi \rho)$. }
\end{figure}

To understand the physical properties of the fourth stop band, we investigated the band gap formation at the third-order $\Gamma$ point relative to the lattice harmonic content through FEM simulations. Deriving the analytical expression for the dispersion relations near the fourth stop band is challenging with multiple higher-order Fourier harmonics, including $\epsilon_{1}\cos(K z)$, $\epsilon_{2}\cos(2K z)$, $\epsilon_{3}\cos(3K z)$, and $\epsilon_{4}\cos(4K z)$. Central to our approach is splitting the contributions corresponding to the individual Fourier harmonics. Figures \ref{fig3}(a)--\ref{fig3}(e) illustrate the FEM-simulated fourth stop bands and $Q$ factors for the 1D lattices with dielectric functions of $\epsilon_{0} + \epsilon_{1}\cos(K z)$, $\epsilon_{0} + \epsilon_{2}\cos(2K z)$, $\epsilon_{0} + \epsilon_{3}\cos(3K z)$, $\epsilon_{0} + \epsilon_{4}\cos(4K z)$, and $\epsilon_{0} + \epsilon_{5}\cos(5K z)$, respectively. As the fourth stop band opens at the third-order $\Gamma$ point ($k_z=2K$ in the extended Brillouin zone) under the Bragg condition $k_z = q(\pi/p)$, in which $p$ is the period of the dielectric constant modulation and $q$ represents the Bragg order, it is reasonable to interpret that the stop bands denoted by $\Delta \Omega_{1}$, $\Delta \Omega_{2}$, and $\Delta \Omega_{4}$ are formed by the fourth-order Bragg process, $\mathrm{BR}_{4,1}$, due to the first Fourier harmonic, the second-order Bragg process, $\mathrm{BR}_{2,2}$, due to the second Fourier harmonic, and the first-order Bragg process, $\mathrm{BR}_{1,4}$, due to the fourth Fourier harmonic, respectively. The third Fourier harmonic with a period of $\Lambda/3$ and the higher-order Fourier harmonics $\epsilon_{n \geq 5}\cos(n K z)$ cannot contribute to the fourth stop band by themselves. As can be seen by Figs. \ref{fig3}(a), \ref{fig3}(b), and \ref{fig3}(d), the fourth band gap is controlled primarily by $\mathrm{BR}_{1,4}$ because $\mathrm{BR}_{2,2}$ and $\mathrm{BR}_{4,1}$ are substantially weaker than the first-order Bragg process. We note that the band structure of the non-approximated BDG with full Fourier harmonics is very close to that of the approximated BDG with $\epsilon_{0} + \epsilon_{4}\cos(4K z)$. Because the fourth Fourier harmonic coefficient $\epsilon_{4}=(\Delta \epsilon/2\pi) \sin (4\pi \rho)$ changes its sign three times from $+$ to $-$, or $-$ to $+$, when $\rho_{m=1,2,3}=0.25\times m$, it is reasonable to conclude that the fourth band gap closes near $\rho_{m=1,2,3}$ when the three major scattering processes, $\mathrm{BR}_{1,4}$, $\mathrm{BR}_{2,2}$, and $\mathrm{BR}_{4,1}$, are balanced destructively. The simulated spatial electric field distributions at the upper and lower bands of $\Delta\Omega_4$, illustrated in Fig. \ref{fig4}, are suffice to demonstrate that band flips occur because the primary first-order scattering process ,$\mathrm{BR}_{1,4}$, tends to locate the symmetric (asymmetric) Bloch modes at the lower (upper) band edges when $0 < \rho < 0.25$ and  $0.50 < \rho < 0.75$. The contribution of the auxiliary processes $\mathrm{BR}_{2,2}$ and $\mathrm{BR}_{4,1}$ is limited to moving the exact locations of the transition points.

\begin{figure}[t]
\includegraphics[width=8.3 cm]{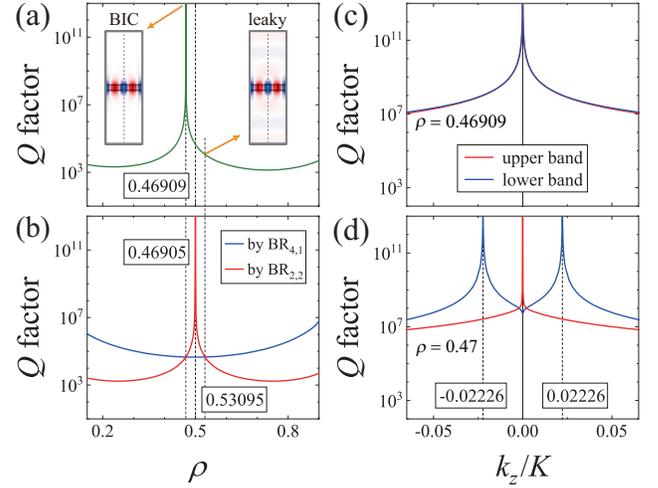}
\caption {\label{fig5} (a) Leaky edge $Q$ factors at the fourth stop band in the non-approximated BDG with full Fourier harmonics as a function of $\rho$. (b) Simulated radiative $Q$ factors of the leaky edge modes in the approximated lattices with dielectric functions $\epsilon_{0}+\epsilon_{1}\cos(K z)$ and $\epsilon_{0}+\epsilon_{2}\cos(2K z)$. Radiative $Q$ factors in the lower and upper band branches in the full lattice as a function of $k_z$ when (c) $\rho = 0.46909$ and (d) $\rho = 0.47$. }
\end{figure}

Now, we demonstrate that a new type of BIC is formed via the destructive interplay between the diffractions by the first and second Fourier harmonic components. The simulated radiative $Q$ factors illustrated in Fig. \ref{fig3} indicate that, in the vicinity of the fourth stop band, out-of-plane radiation loss is caused primarily by the second-order $\mathrm{BR}_{2,2}$ due to the second Fourier harmonic, and to a lesser extent by the fourth-order $\mathrm{BR}_{4,1}$ due to the first Fourier harmonic. Higher-order Fourier harmonics $\epsilon_{n\geq 3}\cos(nK z)$ do not contribute to the radiative $Q$ factors by themselves. Even though the first-order $\mathrm{BR}_{1,4}$ process caused by the fourth Fourier harmonic contributes considerably to opening the fourth band gap, the out-of-plane radiation due to the first-order Bragg process is suppressed by the total internal reflection at the fourth stop band. As shown in Fig.~\ref{fig2}(h), the second band gap can be closed because of the destructive interaction between the first and second Fourier harmonics owing to the variation of $\rho$ \cite{SGLee2019-1,SGLee2019-2}. Similarly, it is reasonable to expect that as $\rho$ varies from 0 to 1, a critical value of $\rho$ exists for which the radiation loss at the fourth stop band vanishes as a result of the destructive interaction between the first and second Fourier harmonics. To verify our conjecture, we investigated the radiative $Q$ factors of leaky edge modes with the frequency $\omega^{s}$ in the BDG as a function of $\rho$ with the results plotted in Fig.~\ref{fig5}(a). Indeed, Fig.~\ref{fig5}(a) demonstrates that there is a critical value ($\rho = 0.46909$) at which the leaky edge mode exhibits a $Q$ value exceeding $10^{14}$ even though it is not protected by the in-plane symmetry. Then, to explore the formation mechanism of the new BIC, the radiative $Q$ factors of leaky edge modes in the approximated lattices with dielectric functions $\epsilon_{0} + \epsilon_{1}\cos(K z)$ and $\epsilon_{0} + \epsilon_{2}\cos(2K z)$ were investigated. As illustrated in Fig.~\ref{fig5}(b), the $Q$ factors corresponding to the $\mathrm{BR}_{4,1}$ ($\mathrm{BR}_{2,2}$) process due to the first (second) Fourier harmonic becomes minimal (diverges to a infinite value) when $\rho =0.5$ because this is where the first (second) Fourier coefficient is maximal (zero). There are two fill factors, $\rho = 0.46905\simeq 0.46909$ and $0.53095$, where the leaky edge $Q$ factor ($\sim5\times 10^4$) due to the first Fourier harmonic is the same as that due to the second Fourier harmonic. As the signs of the first and second Fourier coefficients are the same (opposite) when $\rho < 0.5$ ($\rho > 0.5$), we conclude herein that the new type of BIC with a radiative $Q$ factor up to $10^{14}$ appears at the third-order $\Gamma$ point because the radiative waves due to the first and second Fourier harmonics interfere with each other destructively. By contrast, when $\rho = 0.53095$, the $Q$ value ($\sim\times 10^4$) for the BDG becomes smaller than that ($\sim5\times 10^4$) due to the first or second Fourier harmonic because the radiative waves interfere constructively. The spatial electric field distributions in the insets in Fig.~\ref{fig5}(a) evidently show that leaky edge mode with $\rho = 0.46909$ is confined within the grating layer, while the leaky edge mode with $\rho = 0.53095$ radiates out of the grating.

The formation of the new type of BICs can also be seen by investigating the radiative $Q$ factors of the Bloch modes as a function of $k_z$. For example, when $\rho = 0.46909$, as illustrated in Fig.~\ref{fig5}(c), both the symmetric and asymmetric edge modes become BICs with $Q$ factors exceeding $10^{14}$ at $k_z = 0$, although the $Q$ values decrease gradually as $k_z$ moves away from the $\Gamma$ point. However, the Bloch modes in both the upper and lower bands have high $Q$ values ($> 10^7$) across the entirety of the simulated range of $|k_z| \leq 0.065~K$. The $Q$ factors in the lattice with $\rho = 0.46909$ (shown in Fig.~\ref{fig5}(c)) are approximately $10^3$ times larger than those with $\rho = 0.35$ (see Fig.~\ref{fig2}(b)). Figure \ref{fig5}(d) also reveals that the new type of BIC at $k_z =0$ is split into two BICs at $k_z\neq 0$. The locations of the BICs increasingly deviates from the $\Gamma$ point as $|\rho - 0.46909|$ increases from zero. It is worth while comparing the enhanced $Q$ factors around the $\Gamma$ point, represented in Fig. \ref{fig5}(c), with the topologically enabled ultrahigh-$Q$ resonances reported in a recent study \cite{JJin2019}. When multiple BICs merge at the $\Gamma$ point, the radiative $Q$ factors near the $\Gamma$ point are strongly enhanced and high $Q$ factors resistant to out-of-plane scattering can be achieved. While the topologically enhanced $Q$ factors reported in the previous study were achieved around the second-order $\Gamma$ point by tuning the period of the lattice, our results show that the high $Q$ values can be obtained around the third-order $\Gamma$ point by varying the fill factor, $\rho$, of the lattice.

\begin{figure}[t]
\includegraphics[width=8.3 cm]{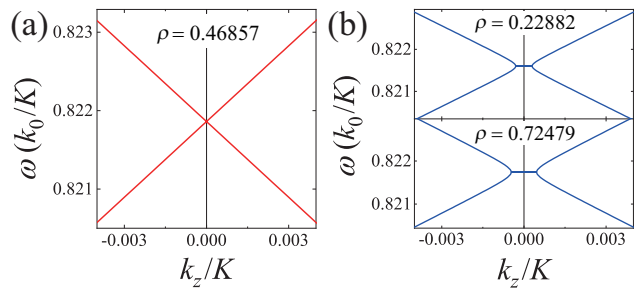}
\caption {\label{fig6} Dispersion relations at the closed band states in the vicinity of the third-order $\Gamma$ point in the BDG structure with (a) $\rho = 0.46857$, (b) $\rho = 0.22882$, and (c) $\rho = 0.72479$. Dirac cone dispersion can be obtained when the out-of-plane radiation loss is suppressed. }
\end{figure}

\begin{figure}[t]
\includegraphics[width=8.3 cm]{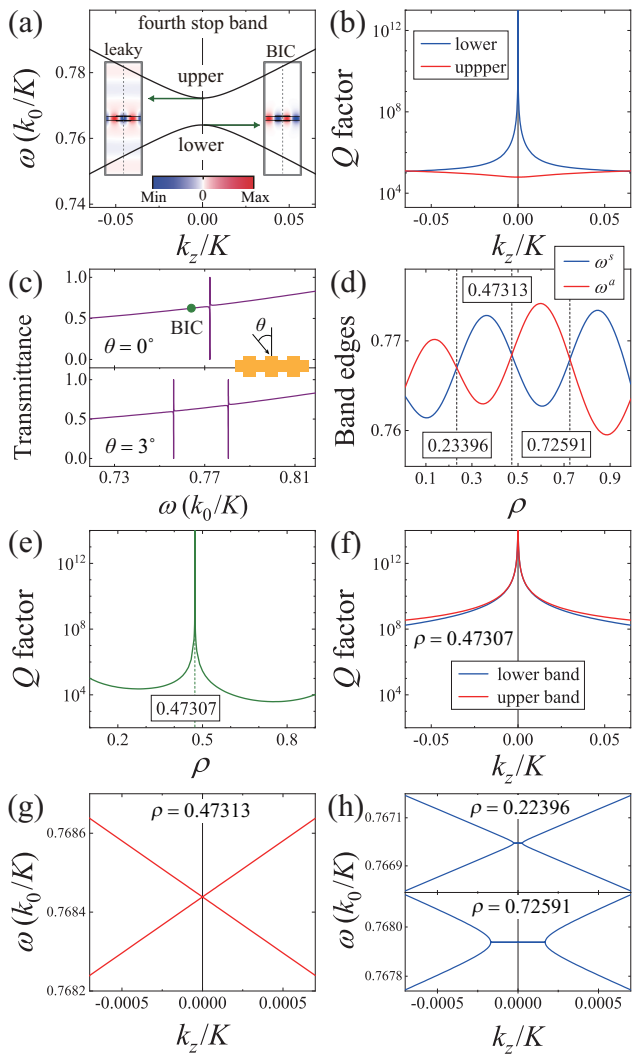}
\caption {\label{fig7} Properties of the DSG structure at the fourth stop band. Dispersion relations (a) and radiative $Q$ factors (b) near the third-order $\Gamma$ point. (c) Transmission spectra exhibiting Fano resonances when $\theta = 0^\circ$ and $\theta = 3^\circ$. (d) Evolution of the band edge frequencies as a function of $\rho$. (e) Leaky edge $Q$ factors as function of $\rho$. (f) Radiative $Q$ factors in the lower and upper band branches as a function of $k_z$ when $\rho = 0.47307$. Dispersion relations at the closed band states when (a) $\rho = 0.47313$, (b) $\rho = 0.22396$, and (c) $\rho = 0.72591$. }
\end{figure}

Next, we analyzed the closed band states at the third-order $\Gamma$ point. As indicated in Figs. \ref{fig2}(d) and \ref{fig5}(c), the value of $\rho = 0.46857$ at which the fourth band gap closes is very close to $\rho = 0.46909$ for which the radiative $Q$ factors are significantly enhanced in the vicinity of the third-order $\Gamma$ point. It has been demonstrated that a closed band state with suppressed out-of-plane scattering loss leads to Dirac cone dispersion, which is currently of great scientific interest \cite{Minkov2018,SGLee2020-2}. Mathematically, Dirac cones refer to the closed band states with crossing dispersion curves as straight lines in the vicinity of the $\Gamma$ point. Physically, Dirac cones represent massless photonic states and are associated with many interesting phenomena, such as photonic topological insulators \cite{Ziolkowski2004,AAlu2007,Liberal2017} and zero-refractive index materials \cite{Haldane2008,LHWu2015,LLu2014}. As shown in Fig. \ref{fig6}(a), at the closed band state with $\rho = 0.46857$, the dispersion curves cross as straight lines, while $\partial \omega/\partial k_{z} \neq 0$ at the third-order $\Gamma$ point. Conversely, at the closed band state with $\rho = 0.22882$ or 0.72479, there are finite ranges of Bloch wave vectors $\Delta k_{z}$, for which $\partial \omega/\partial k_{z} = 0$, that ruin the Dirac cone dispersion.

Band dynamics accompanied by the BIC at the third-order $\Gamma$ point in the DSG structure were also investigated, with the fundamental properties summarized in Fig. \ref{fig7}. The fourth band gap opens at the third-order $\Gamma$ point, as shown in Fig. \ref{fig7}(a). The spatial electric field distributions shown in the insets in Fig. \ref{fig7}(a) indicate that one of the band edge modes with an asymmetric field distribution becomes the nonleaky symmetry-protected BIC, while the other one, with a symmetric field distribution, radiates out of the lattice. As can be seen by Fig. \ref{fig7}(b), the symmetry-protected BIC in the lower band exhibits a $Q$ factor that is larger than $10^{14}$ at the $\Gamma$ point, but the $Q$ values decrease abruptly and tend toward the $Q$ factor in the upper band ($\sim 10^5$) as $k_z$ moves away from the $\Gamma$ point. Figure. \ref{fig7}(c) demonstrates that, at normal incidence with $\theta = 0^\circ$, only one resonance, caused by the leaky mode in the upper band, is observed. The symmetry-protected BIC in the lower band is not shown in the transmittance curve. By contrast, when $\theta = 3^\circ$, tworesonances that can be attributed to the Bloch modes in the upper and lower bands appear concurrently in the transmission spectra. As $\rho$ varies from zero to 1, the fourth stop band exhibits three closed band states when $\rho = 0.23396$, 0.47313, and 0.72591 as shown in Fig. \ref{fig7}(d). Before and after the band gap closures, the relative position of the leaky edge BIC edges are reversed. Figure \ref{fig7}(e) evidently shows that a specific value of $\rho = 0.47307$ exists for which the leaky edge $Q$ factor diverges to infinity. As revealed in Fig. \ref{fig7}(f), in the DSG with $\rho = 0.47307$, the $Q$ factors in the lower and upper bands are significantly enhanced in the entirety of the simulated range of $|k_z| \leq 0.065~K$. Furthermore, Fig. \ref{fig7}(g) also demonstrates that, at the closed band state with $\rho = 0.47313$, the dispersion curves cross as straight lines and $\partial \omega/\partial k_{z} \neq 0$ at the third-order $\Gamma$ point, which is attributed to the strong suppression of out-of-plane scattering losses by the destructive interplay between the first and second Fourier harmonics. Alternatively, at the closed band states with $\rho = 0.23396$ and 0.72591, Fig. \ref{fig7}(h) clearly shows that Dirac cone dispersions are ruined owing to the finite ranges of Bloch wave vectors $\Delta k_{z}$ for which $\partial \omega/\partial k_{z} = 0$ \cite{SGLee2019-1,SGLee2020-2}. To summarize, the key properties of the DGS structure, illustrated in Fig. \ref{fig7}, are the same as those of the BDG.

\begin{figure}[t]
\includegraphics[width=8.3 cm]{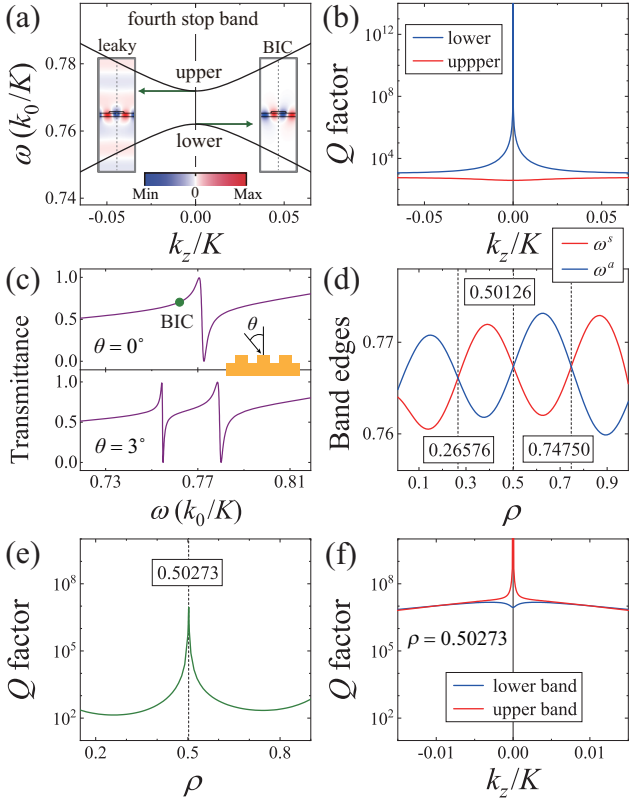}
\caption {\label{fig8} Properties of the OSG structure at the fourth stop band. Dispersion relations (a) and radiative $Q$ factors (b) near the third-order $\Gamma$ point. (c) Transmission spectra that exhibiting Fano resonances when $\theta = 0^\circ$ and $\theta = 3^\circ$. (d) Evolution of the band edge frequencies as a function of $\rho$. (e) Leaky edge $Q$ factors as function of $\rho$. (f) Radiative $Q$ factors in the lower and upper band branches as a function of $k_z$ when $\rho = 0.50273$. }
\end{figure}

We also considered the band dynamics at the fourth stop band in the OSG structure. For practical applications, fabrication of the OSG is more convenient than that of DSG with up-down mirror symmetry. As can be seen by Figs. \ref{fig8}(a)--\ref{fig8}(d), the dispersion relations, radiative $Q$ factors, transmission spectra, and evolution of band edge frequencies for the OSG structure exhibit the same tendencies as in the DSG and BDG structures. However, a noticeable difference between the OSG and DSG structures can be found by comparing the simulated $Q$ factors as a function of $\rho$ from Fig. \ref{fig8}(e) and Fig. \ref{fig7}(e). While the $Q$ values for the DSG seem to diverge to an infinite value at $\rho = 0.47307$, the $Q$ factors for the OSG (Fig. \ref{fig8}(e)) are saturated to a finite value ($\sim 10^7$) at $\rho = 0.50273$. By inspecting Figs. \ref{fig5}, \ref{fig7}, and \ref{fig8}, it is reasonable to conclude that the new type of BIC, which is caused by the destructive interplay between the first and second Fourier harmonics, can be achieved in photonic lattices with up-down mirror symmetry. This conclusion is consistent with previous studies in which various BICs beyond the protection of in-plane symmetry have been introduced \cite{XGao2016,LNi2016,SGLee2020-1}. However, Fig. \ref{fig8}(f) evidently shows that $Q$ factors in both the lower and upper bands are notably increased in the vicinity of the $\Gamma$ point when $\rho = 0.50273$. The $Q$ factors in the lattice for which $\rho = 0.50273$ are approximately $10^3$ times larger than those with $\rho = 0.40$ (see Fig. \ref{fig8}(b)).

In this study, we completed a detailed analysis of the fourth stop bands of leaky-mode photonic lattices. Our results indicate that the fourth stop band can potentially be used to manipulate electromagnetic waves, similar to how second stop bands have been exploited. One advantage of the utilizing the fourth stop band may be the convenience of the nanofabrication. Since the photonic band gaps are open at the Bragg condition $k_z = qK/2$, the pitch size of a photonic device employing the fourth stop band can be approximately two times larger than that employing the second stop band. Furthermore, Dirac cone dispersions can be realized at the closed band state owing to the enhanced $Q$ factors in the vicinity of the third-order $\Gamma$ point.

%\section{Conclusion}
In conclusion, we investigated the band dynamics accompanying BICs at the third-order $\Gamma$ point in three representative 1D photonic lattices through rigorous FEM simulations. Our analyses show that the fourth stop band is controlled primarily by the first-order diffraction arising from the fourth Fourier harmonic lattice component, and to a lesser extent by the second-order and fourth-order diffractions caused by the second and first Fourier harmonic contents, respectively. Near the fill factors of 0.25, 0.50, and 0.75, for which the fourth-order Fourier coefficient becomes zero, the auxiliary processes become competitive with the primary process. It is the interplay between these three major processes that closes the fourth band gap and induces the band flip whereby the leaky and BIC edges transit across the fourth band gap. It was also revealed that the out-of-plane radiation loss at the fourth stop band is caused primarily by the second Fourier harmonic component, and to a lesser degree by the first Fourier harmonic. In the BDG and DSG lattices with up-down mirror symmetry, new type of BICs are formed because of the destructive interference between the second-order diffraction by the second Fourier harmonic and the fourth-order diffraction by the first Fourier harmonic. When the fourth band gap closes near a fill factor of 0.5, Dirac cone dispersions can be formed because the out-of-plane radiation is significantly suppressed by the destructive interplay between the first and second Fourier harmonics. In the OSG, which lacks the up-down mirror symmetry, $Q$ factors at the fourth stop band also increase noticeably owing to the destructive interplay between the first and second Fourier harmonics. However, a new type of quasi-BIC are observed in the OSG because the $Q$ values are saturated to a finite value. Although this study was restricted to the band dynamics of the lowest fundamental mode in 1D leaky-mode photonic lattices, the extension of this work to higher-order guided modes and 2D lattices is feasible. Moreover, this study may be helpful in manipulating electromagnetic waves by utilizing high $Q$ Bloch modes near the third-order $\Gamma$ point.

This research was supported by grants from the National Research Foundation of Korea, funded by the Ministry of Education (No. 2020R1I1A1A01073945) and Ministry of Science and ICT (No. 2020R1F1A1050227), along with the Gwangju Institute of Science and Technology Research Institute in 2020.

\end{document}